# AIxArtist: A First-Person Tale of Interacting with Artificial Intelligence to Escape Creative Block

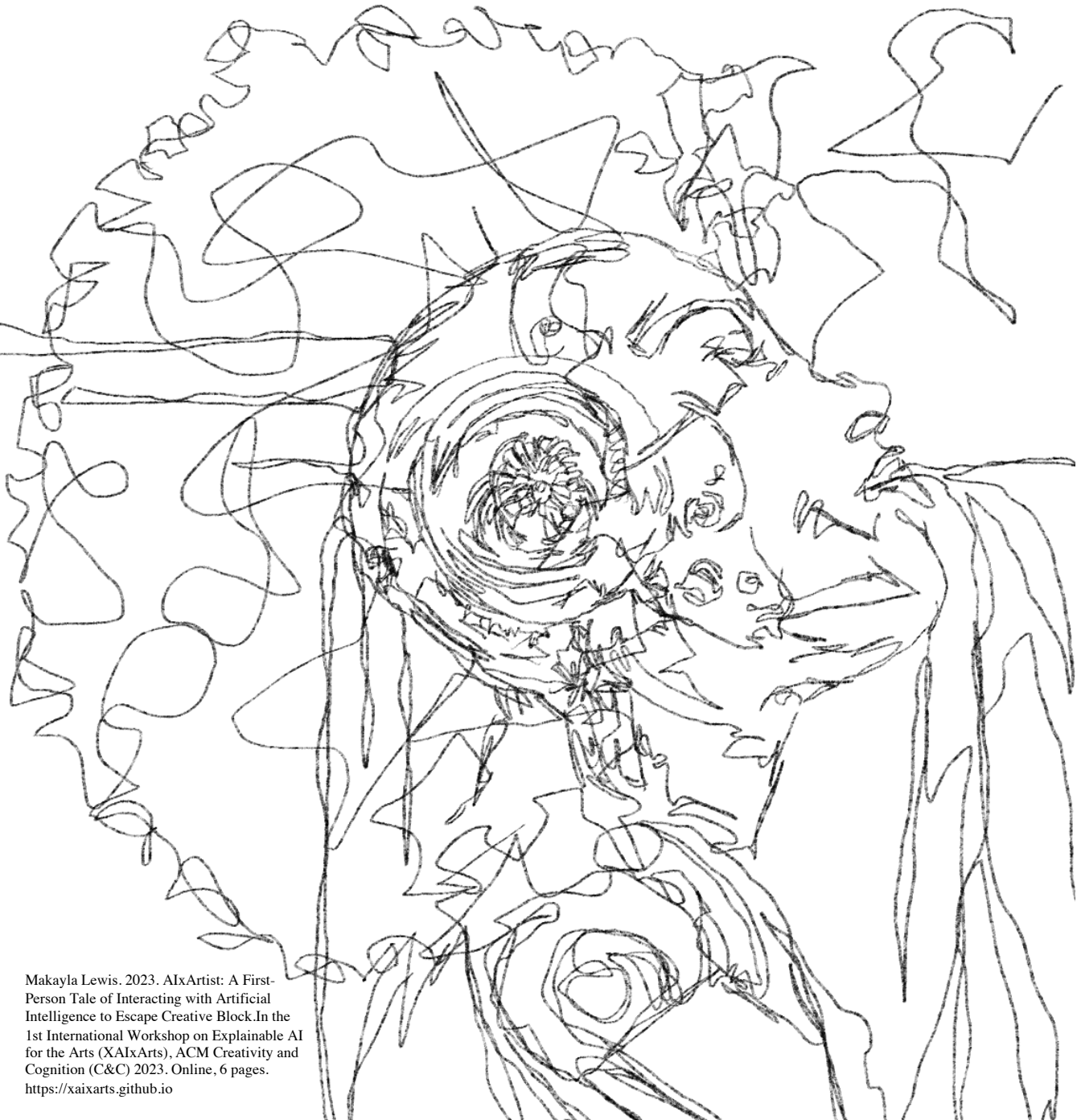

**Makayla Lewis**
Digital Media Kingston, Kingston University
London, United Kingdom
m.m.lewis@Kingston.ac.uk

**ABSTRACT**
The future of the arts and artificial intelligence (AI) is promising as technology advances. As the use of AI in design becomes more widespread, art practice may not be a human-only art form and could instead become a digitally integrated experience. With enhanced creativity and collaboration, artists and AI could work together towards creating artistic outputs that are visually appealing and meet the needs of the artist and viewer. While it is uncertain how far the integration will go, arts and AI will likely influence one another. This workshop pictorial puts forward first-person research that shares interactions between an HCI researcher and AI as they try to escape the creative block. The pictorial paper explores two questions: How can AIs support artists' creativity, and what does it mean to be explainable in this context? AIs, *ChatGPT* and *MidJourney* were engaged. The result was a series of reflections that require further discussions/explorations in the XAIxArts community: Transparency of attribution, the creation process, ethics of asking, and inspiration vs copying.

**Author Keywords**
Arts; Artificial Intelligence; Creative Collaboration; Creative Block; First Person Research.

**CSS Concepts**
• Human-centered computing ~ Human computer interaction (HCI) ~ Empirical studies in HCI.

**Figure 1:** AI and Me, 2023 "May be beyond?". Digital sketchbook entry, a response to ChatGPT prompt 4 and 7 (see Figure 4 middle) and MidJourney `/imagine` inspirational outputs (see Figure 7). Apple Pencil on iPad Pro 11 using Procreate by Makayla Lewis, 2023.

**INTRODUCTION**

Art has been an integral part of human civilisation for thousands of years. People have created and expressed themselves differently throughout history, from the earliest cave paintings to contemporary digital art [4, 8, 10]. Art refers to all forms of creative expression, including drawing, sketching, painting, sculpture, photography, printmaking, and many others, e.g. [13, 17, 18]. Art practice is often used as a communication, documentation, and expression tool. With the advent of digital technology, artists can now create digitally and share online, making their work accessible to a global audience [7]. Today, digital tools have become essential to art practice (visualising ideas and emotions and documenting the world around us).

**Arts and HCI**

In Human-Computer Interaction (HCI), art can convey design ideas and concepts, e.g. [2, 3, 7, 8, 9, 10, 16, 18]. It can effectively visualise and communicate design concepts, experiences, and emotions quickly and clearly, allowing creators to explore the digital differently. It can also generate user feedback during the design process by engaging with participatory design and co-design, e.g. [1, 13].

In recent years, there has been greater interest in arts in computer science, primarily driven by the rise of machine learning and artificial intelligence (AI). HCI researchers are now beginning to look at the use of AI to explore how the integration of arts in HCI can be improved.

Although there is an increasing interest in Arts and HCI, the role of the 'researcher as artist/artist as a researcher [18] is few in digital media and HCI departments. As such, creativity support can be limited: the need for *encouragement and validation* of one's efforts, the creative and reflective *space and time* to allow ideas to develop, resources that *inspire and support* creative flow, and the ability to *collaborate* creatively are lacking (in some instances, this is missing). As a result, a reduction in creativity can occur, impacting not only their art skill and practice but also their professional work.

**CASE STUDY**

Art block, also known as creative block, is a phenomenon that can occur when an artist experiences a lack of inspiration or motivation to create new work. It can affect artists of all levels, from beginners to professionals. It can be attributed to many reasons, e.g., imposter syndrome, stress, lack of variety in routine, and being overwhelmed or burnout [16]. Makayla Lewis: *"Semester two, mid-January to end to March, is always stressful and overwhelming; teaching, supervising projects, publication deadlines, and funding bids mean I often experience creative block. A mental block where ideas disappear; they do not flow like they usually do. It is frustrating to look at one's creative tools, touch them and do nothing in favour of less creative/demanding pursuits. It often lasts the length of the semester and can negatively impact one's working practices and relationships. Furthermore, coming out of the creative block at the end of the semester is more challenging than being in it, the fear of failure being the core driver of struggling to escape".*

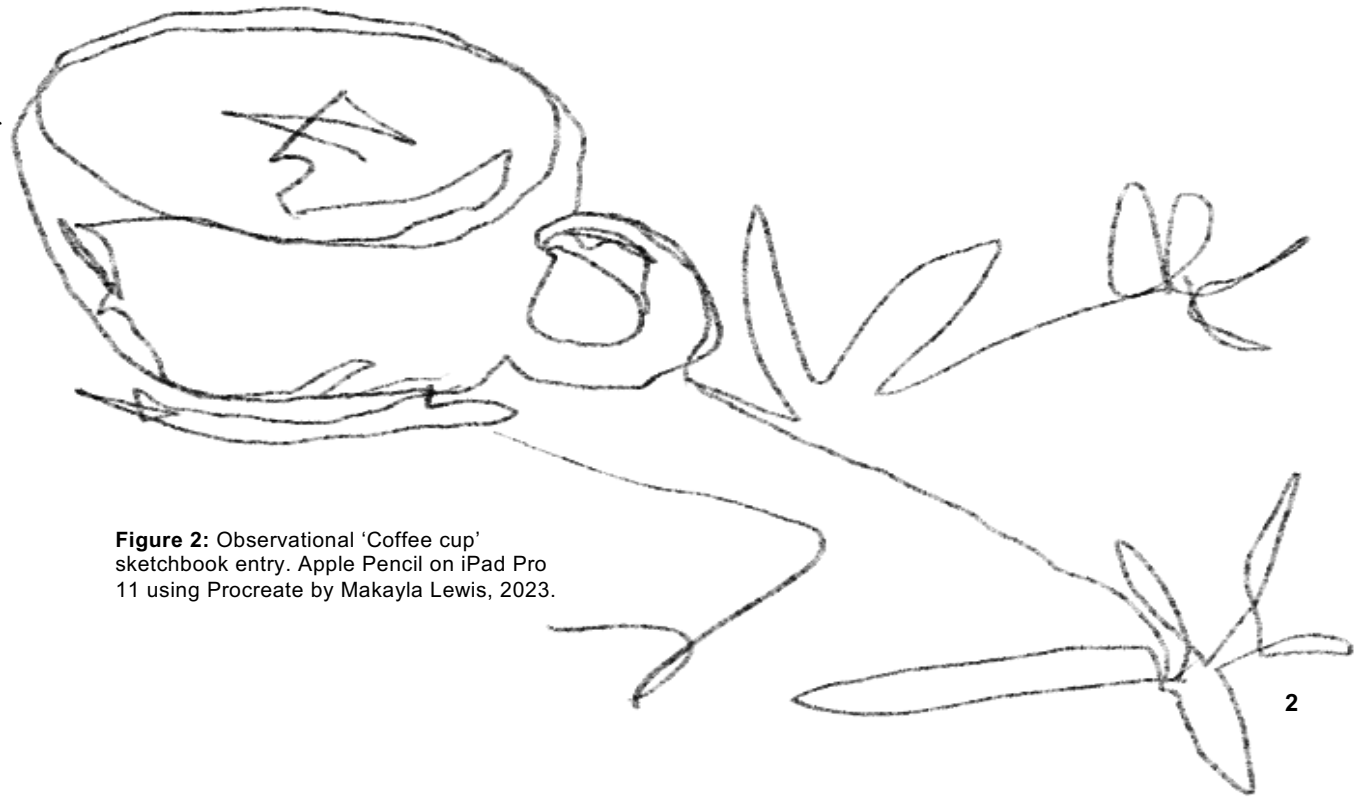

**Figure 2:** Observational 'Coffee cup' sketchbook entry. Apple Pencil on iPad Pro 11 using Procreate by Makayla Lewis, 2023.

This exploratory workshop pictorial takes a first-person research method [5, 14] to present recent experiences, reflections, and example artworks of an HCI researcher who attempts to escape art block through engagement with AI: *Chat GPT* and *MidJourney*.

**Tale of Creativity Renewal**

The decision to escape this year's creative block started during the Easter Holiday, post-semester. It began by retrieving art practice tools and attempting to draw the first thing that came to mind; nothing flowed, and the outputs were limited and disjointed (e.g., Figures 1 and 2). Thus, most blank pages became all-encompassing and were soon abandoned to browse Twitter. Makayla Lewis: *"As an artist and researcher, I followed many artists on Twitter and 'lurked' discussions concerning the role of AI and arts; some artists thought it could be beneficial, whereas others were not convinced and, in some instances, were highly negative"*—most of the conversations centred around AI art generators and the role of *ChatGPT* in the arts.

This interest was further compounded by a recent conversational design workshop that Makayla Lewis attended, which focused on interaction with *MidJourney*. An AI image generator that offers users a gothic style upon entering `/imagine` prompts. The vibrant social media community discussions and experiences at the workshop sparked a greater intrigue: **Could *ChatGPT* and *MidJourney* support the escape of creative block?** As a result, the journey to renew creativity began.

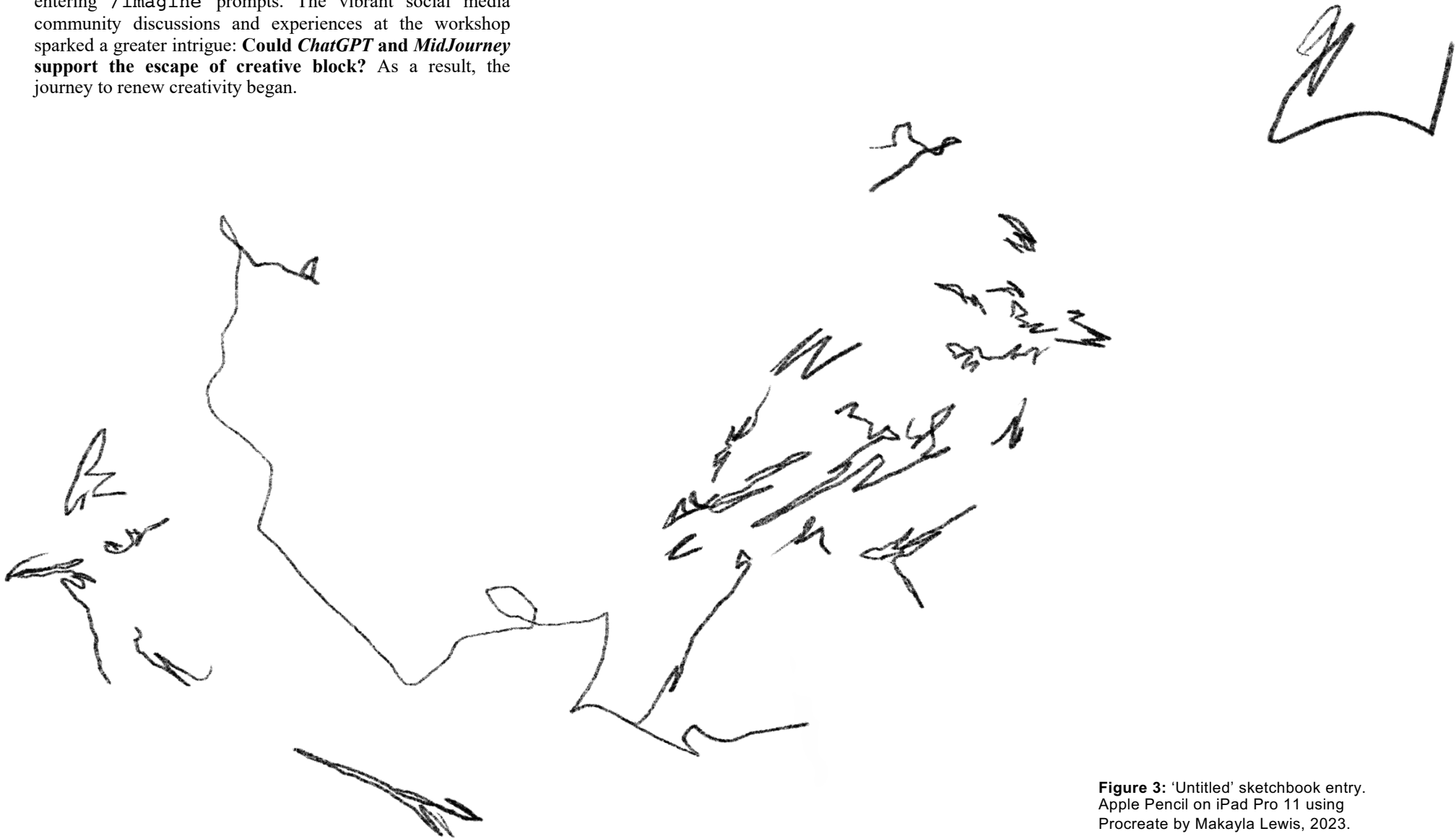

**Figure 3:** 'Untitled' sketchbook entry. Apple Pencil on iPad Pro 11 using Procreate by Makayla Lewis, 2023.

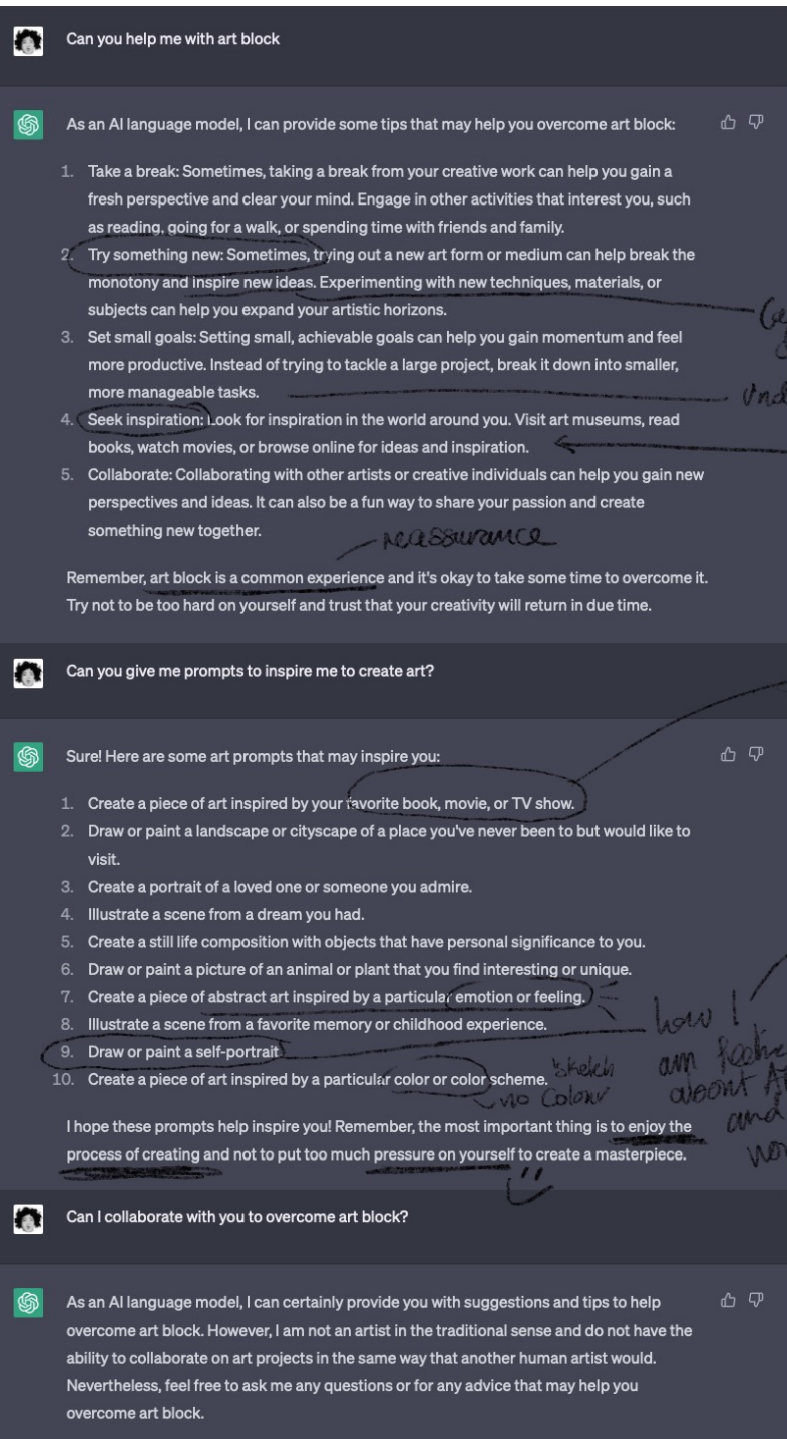

*ChatGPT* responded to three questions asked by Makayla Lewis (Figure 4-6 left); the responses were exported to *Procreate App,* then annotated, and the art piece was brainstormed (Figure 7). *ChatGPT* Art Block prompts 1, 7 and 9 leapt out as possibilities. Makayla Lewis: "*Sketching oneself is always easy. However, incorporating a science fiction element was new yet sounded fun*". Thus, a How Might We question was identified: **What role is AI playing in my creative block escape?** As an artist not akin to sketching science fiction, *MidJourney* was visited for visual inspiration using prompts, e.g., artist and AI, creativity and AI. (Figure 7 below). The outcome was a sketchy reflective self-portrait 'AI and Me, 2023 "Maybe beyond?" (Figure 1 on page 1). This process was created in 20 minutes and, as a result, launched ML into personal work. Makayla Lewis' art practice did not cease after this engagement with AIs; they have since walked away from *MidJourney* to create personal works through sketches, digital paintings, and sketchnotes in their environment, e.g., Figures 8 and 9. These were influenced by *ChatGPT* prompt 5 and 6 (Figure 4-6 left).

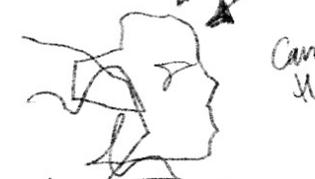
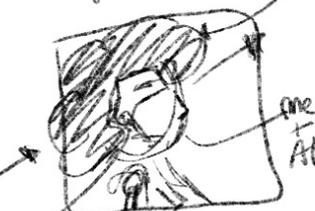
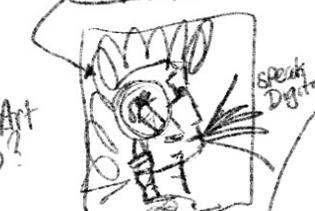
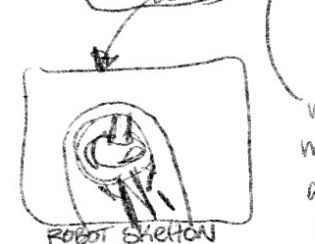
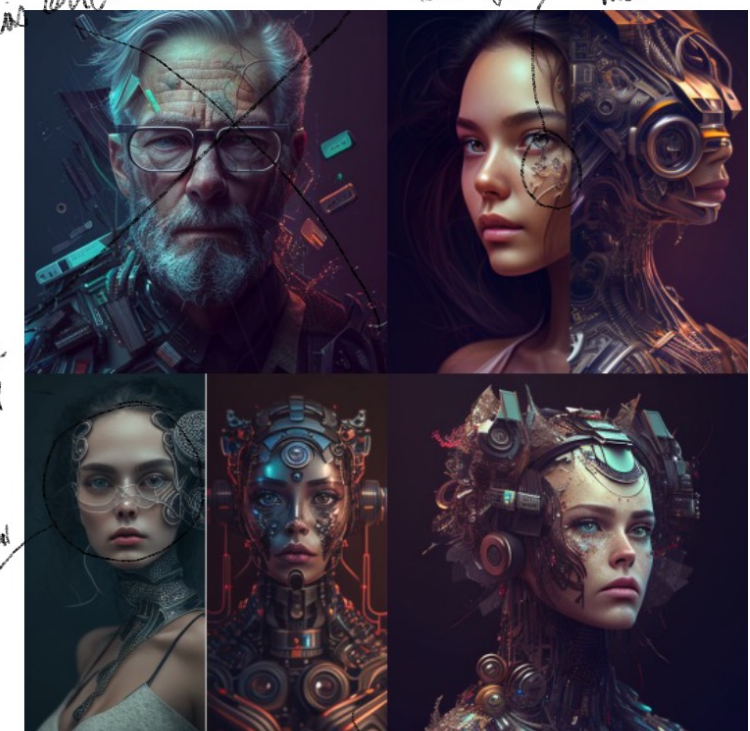

**Figure 4-6 (left):** Screenshots of ChatGPT responses.

**Figure 7 (above):** Screenshots /imagine prompts to inspire— annotations and sketches by Makayla Lewis, 2023.

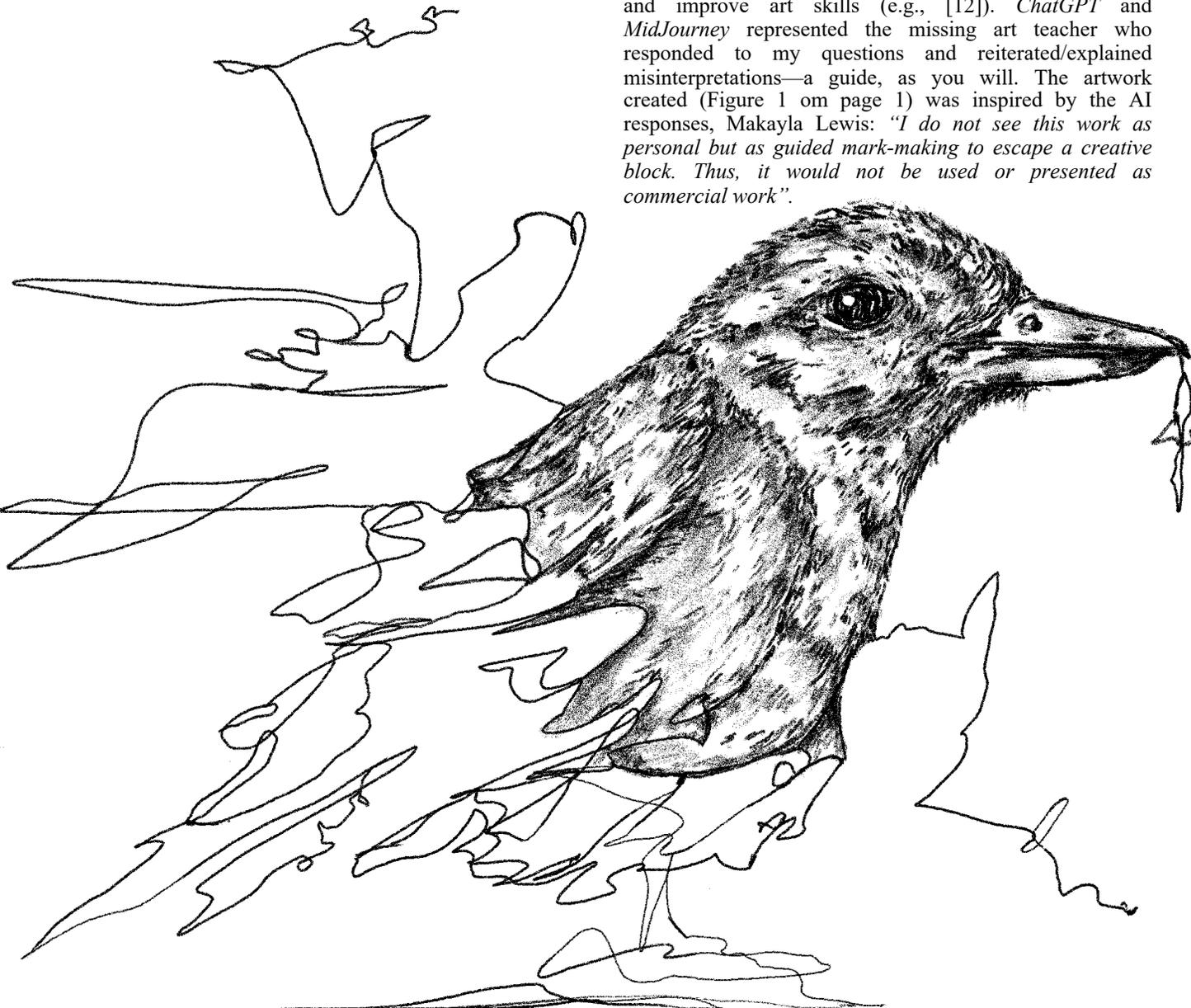

**Figure 8:** 'Appearance of Warbling Vireo' digital sketchbook entry. Kindle Scribe using Notebook by Makayla Lewis, 2023.

## REFLECTIONS

Although not perfect, the process was well practised; it was a remembrance of school days when an art teacher would give a prompt and share examples of well-known art masters. In which the student was inspired to create work in class or for an assignment. Its purpose of studying Master's and improve art skills (e.g., [12]). *ChatGPT* and *MidJourney* represented the missing art teacher who responded to my questions and reiterated/explained misinterpretations—a guide, as you will. The artwork created (Figure 1 om page 1) was inspired by the AI responses, Makayla Lewis: *"I do not see this work as personal but as guided mark-making to escape a creative block. Thus, it would not be used or presented as commercial work"*.

The process brought to light four concerns that require more significant discussion and exploration in the XAIxArts community: **transparency of attribution,** AI's views contributions of artists across the internet (studying the work of different artists), but who are the contributors? why were they selected? Were permissions obtained? and does this lack of transparency impact trust in the responses and or negatively impact the arts community?; **the creation process,** studying others (human or AI) requires a clear understanding of the creation process, how and why were responses constructed in such a way? **ethics of asking,** is a quick fix, non-human engagement, to escape creative block appropriate? Would engagement with the physical world be more beneficial to boosting creativity?; and **inspiration vs copying** art practice and personal art production have different roles in artists' creative process; ensuring they remain separate is vital to a thriving arts community. Thus, how will AI's art prompts and visual generators ensure separation?

This workshop position pictorial does not seek to answer or respond to these reflections but instead put them forward to the XAIxArts communities, with an exemplar, for further discussion and exploration.

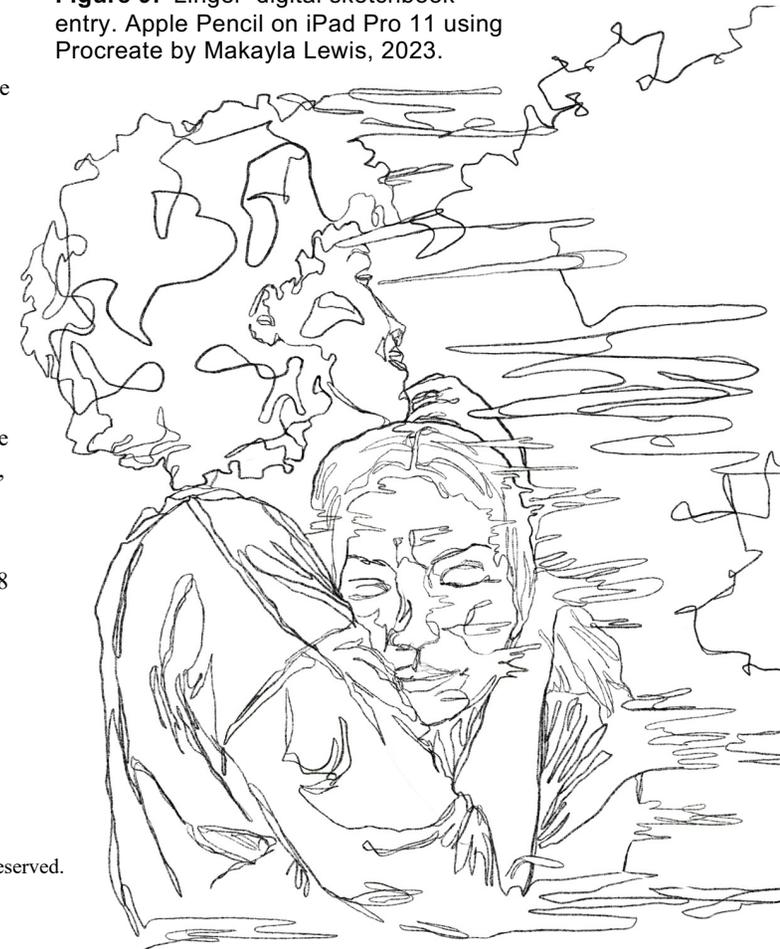

**Figure 9:** 'Linger' digital sketchbook entry. Apple Pencil on iPad Pro 11 using Procreate by Makayla Lewis, 2023.